\documentclass[aps,
               prl,
               twocolumn,
               10pt,
               longbibliography,
               floatfix,
               nofootinbib,
               superscriptaddress]{revtex4-2}\usepackage[T1]{fontenc}
\usepackage{amsmath}
\usepackage[dvipsnames]{xcolor}
\usepackage{physics}
\usepackage{graphicx}
\usepackage{dcolumn}
\usepackage{bm}
\usepackage{hyperref}
\usepackage{array}
\usepackage{etoolbox}
\usepackage{graphicx}
\DeclareUnicodeCharacter{00A0}{}
\usepackage{lipsum}
\usepackage{xr}

\usepackage[normalem]{ulem}
\usepackage{amsfonts}
\usepackage{amsmath}
\usepackage{mathtools}
\usepackage{physics}
\usepackage{subfigure}

\usepackage{subfiles}

\emergencystretch=\maxdimen
\allowdisplaybreaks

\makeatletter
\patchcmd{\@makecaption}
  {\scshape}
  {}
  {}
  {}
\makeatother

\usepackage{xr}

\begin{document}



\title{Microscopic Quantum Friction}

\author{Pedro H. Pereira}
\email{pedro_h@id.uff.br}
    \affiliation{Instituto de Física, Universidade Federal Fluminense, 24210-346, RJ, Brazil}

\author{F. Impens}
\author{C. Farina}
\author{P. A. Maia Neto}
\affiliation{Instituto de F\'{i}sica, Universidade Federal do Rio de Janeiro, 21941-972, Rio de Janeiro, Rio de Janeiro,  Brazil}

\author{R. de Melo e Souza}
\email{reinaldos@id.uff.br}
    \affiliation{Instituto de Física, Universidade Federal Fluminense, 24210-346, RJ, Brazil}



\begin{abstract}
We report on a microscopic theory of quantum friction. Our approach investigates the interplay between the dispersive response and the relative center-of-mass motion of two ground-state atoms. This coupling yields a  quantum force, which can be expressed as a power series in the velocity. The significance of each contribution depends on its order parity: while even-order terms are reversible, odd-order terms are irreversible and only survive in the presence of an internal dissipation mechanism. In addition, we obtain general, model-independent properties for the work performed by these contributions for arbitrary scattering trajectories. These results
enable an unambiguous identification of odd-parity terms with microscopic quantum friction. At room temperature,  the dominant microscopic quantum friction is of first order in the velocity and presents a strong quantum character. Our microscopic theory reveals that several properties of quantum friction obtained in specific settings -- such as the cubic dependence on velocity at zero temperature -- are indeed universal features already present at the atomic scale.
\end{abstract}

\maketitle

One of the most elusive effects of the quantum vacuum is quantum friction, a dissipative force acting on matter in relative motion in vacuum. Optically levitated nanoparticles rotating in the gigahertz (GHz) frequency range provides a promising platform for experimental observation of quantum friction~\cite{Manjavacas2010,Xu2021} as well as of alternative non-equilibrium quantum vacuum effects~\cite{Sanders2019,Matos2020,Amaral2025}. Recently, a direct detection was possible with nanomechanical oscillators~\cite{Xu2024},  and quantum friction has been found to play a key role in different material systems in the nanoscales, thus opening a new era for dissipative quantum electrodynamics~\cite{Kavokine2022,Ho2025,Shaukat2025,Franca2025,Gao2025}.
Over the years, quantum friction has sparked continuous interest and conflicting theoretical results~\cite{fact_or_fiction,no_friction,friction_experimental,Milton2016,MILTON2025130475,Reiche2017,Lombardo2017,Farias2018,Farias2019,Viotti2019,Farias2020,Reiche2020,Reiche2022,Fernandez2024,Oue2024,Oue2025}. The core of the controversy 
lies on the intricate treatment of dissipation and non-equilibrium quantum field fluctuations in the framework of the macroscopic Maxwell equations. 
Clearly, a better understanding of the physical mechanism responsible for quantum friction can be 
built at the more fundamental microscopic level. Here we develop a microscopic theory for quantum friction analyzing the dynamical corrections to the van-der-Waals (vdW) interaction between two atoms in arbitrary relative motion.

In contrast to previous works addressing dynamical corrections through electrodynamic retardation effects~\cite{Buhmann09,Donaire2016,Fosco2021,Lombardo2021,Guo2021,Guo2022,Impens13a}, we focus instead on the interplay between the motion and the atomic response, which is primarily responsible for the dynamical corrections at short distances. With this perspective, our treatment unveils several universal features of quantum friction already present at the atomic scale.  For realistic, non-relativistic speeds, we show that it is possible to analyze separately the contribution to quantum friction due to the relative center-of-mass~(CM) motion and the one due to the internal atomic structure. This separation allows us to establish general theorems on the work performed by dynamical vdW corrections for arbitrary motions.~These theorems enable an unambiguous identification of the microscopic quantum friction terms. In particular, we show that the dominant contribution to quantum friction under usual experimental conditions is strongly temperature-dependent.

For static atoms, the dominant contribution to the vdW interaction arises from the UV region of the fluctuation atomic spectrum~\cite{IsraelachviliBook}. Here, the opposite holds for the dynamical corrections to the vdW force, mainly determined by the characteristic internal fluctuations in the low frequency domain.
For transitions with frequency $\omega_0$ the effect of electrodynamic retardation on dynamical corrections can be neglected as long as $\omega_0r/c\ll 1$~\cite{SM}, with $r$ denoting the inter-atomic distance. This condition is easily met for typical values of $r$ and $\omega_0$ and we thus evaluate the non-retarded regime.

\textit{Fluctuation-Polarizability formula for the force---} We first provide a general expression for the dispersion force between moving atoms within  linear response theory. Let us consider two ground-state, unpolarized atoms ($A$ and $B$) set in relative motion. Their CM motion is treated classically. 
The considered distances (of the order of $10\, {\rm nm})$ enable a treatment of the internal atomic charge distribution in the dipole approximation. In the non-retarded regime, the dispersive force stems from the electrostatic field produced by the quantum dipole operators $\hat{\boldsymbol{d}}_A,\hat{\boldsymbol{d}}_B$:
\begin{equation}
    \boldsymbol{F}^B(t)= \langle \boldsymbol{\hat{d}}^B(t)\cdot\boldsymbol{\nabla} \hat{\boldsymbol{E}}^{\rm dip,A} (\boldsymbol{r}(t),t)\rangle\, \label{defforca}
\end{equation}
\begin{equation} \label{E dip}
    \hat{E}_i^{\rm dip,A}(\boldsymbol{r},t)\!=\!\frac{3\hat{r}_i\hat{r}_l-\delta_{il}}{4\pi\varepsilon_0r^3}\hat{d}_l^A(t)\equiv G_{il}(\boldsymbol{r})\hat{d}_l^A(t)\,,
\end{equation}
where $\boldsymbol{r}(t)$ thus denotes the relative position of the CM of $B$ with respect to $A$. 
We have used Einstein's convention and taken the expectation value $\langle ... \rangle$ on the ground states for $A$ and $B$. The  Van der Waals force  on particle $B$ depends on the dipole correlation functions, which can be evaluated within linear response. One obtains~\cite{SM}:
\begin{eqnarray}
       && F^B_i=\frac{\hbar}{2} \left. \partial_mG_{il}(\boldsymbol{r})\right|_{\boldsymbol{r}=\boldsymbol{r}(t)}\int_0^t dt' G_{kj}(\boldsymbol{r} (t'))\times\cr\cr
    &&(\alpha_{mj}^B(t-t')\eta^A_{lk}(t-t')+\alpha_{lk}^A(t-t')\eta^B_{mj}(t-t')) \, , \label{forcacompleta}
\end{eqnarray}
We have introduced the polarizability and fluctuation tensors for each atom $(\xi=A,B)$, defined as $\alpha_{mj}^\mathcal{\xi}(t-t')=\frac{i}{\hbar}\theta(t-t')\langle[\hat{d}^{(0)\xi}_m(t),\hat{d}^{(0)\xi}_j(t')]\rangle$ and $\eta_{kl}^\mathcal{\xi}(t-t')=\frac{1}{\hbar}\langle\{\hat{d}^{(0)\xi}_k(t),\hat{d}^{(0)\xi}_l(t')\}\rangle$, with $\hat{\boldsymbol{d}}^{(0)\xi}(t)$ denoting the free Heisenberg-evolved dipole operator~$(\langle\hat{\boldsymbol{d}}^{(0)\xi}\rangle=0$). From now on, we consider isotropic quantum correlations $\alpha^\mathcal{\xi}_{lk}(t)=\alpha^\mathcal{\xi}(t) \delta_{lk}$ and $\eta_{lk}^\mathcal{\xi}(t)=\eta^\mathcal{\xi}(t) \delta_{lk}$. Dynamical effects are captured by the Green's function $G_{kj}(\boldsymbol{r} (t'))$ evaluated at a time-dependent position. This time-dependence induces an interplay between the  CM motion and the quantum dipole correlations.  
Eq. (\ref{forcacompleta})  highlights the physical intuition of the dispersion interaction: it comes from the superposition of two forces: one between the fluctuating dipole of particle $A$ and the dipole induced in $B$, which is proportional to the polarizability of $B$, and the other with $A$ and $B$ reversing their roles. 

\textit{Dispersive dynamical corrections---} We now work out explicitly the dynamical vdW force corrections~between  moving atoms. Because the interaction is dispersive, there is a typical delay $\tau_m$ in the atomic response, corresponding to the interval on which the polarizability tensor $\alpha_{ij}(t-t')$ assumes significant values. Thus, the dominant contribution to the time-integral in Eq.~\eqref{forcacompleta} arises from  the interval $[t-\tau_m,t]$. Dynamical effects then correspond to a relative motion of the atoms on a distance scale $\delta r \sim |\boldsymbol{\dot{r}}(t)| \tau_m$. A perturbative treatment is appropriate if $\delta r \ll r.$ For the typical response times $\tau_m \sim 1 \: {\rm ps},$ and for distances $r\sim 10 {\rm nm}$, this corresponds to velocities $v \ll 10^{4}$ m/s --- a condition largely fulfilled in realistic experimental conditions.  One can then use a Taylor expansion of the Green function $G_{kj}(\boldsymbol{r}(t'))$ in powers of $\tau=t'-t$, to write the vdW force as a series:  
\begin{equation} \label{serieforca}
 \boldsymbol{F}^B(\boldsymbol{r}(t))=\sum_{n=0}^{\infty} \boldsymbol{D}^{(n)}(\boldsymbol{r}(t))\Lambda^{(n)} \, , 
\end{equation}
with  the  $n$-th order components given by $ \boldsymbol{F}^{(n)}(\boldsymbol{r}(t)) = \boldsymbol{D}^{(n)}(\boldsymbol{r}(t))\Lambda^{(n)}$ where
\begin{eqnarray}
   & & \boldsymbol{D}^{(n)}(\boldsymbol{r}(t))
   = \frac{(-1)^{n}\hbar}{2n!}\left. \boldsymbol{\nabla} G_{kj}(\boldsymbol{r})\right|_{\boldsymbol{r}=\boldsymbol{r}(t)}\frac{d^{n}G_{kj}(\boldsymbol{r}(t))}{dt^n}\,, \label{Disot} \\ 
  & & \Lambda^{(n)}  =  \frac{(-i)^n}{2\pi }\int_{-\infty}^{\infty}
    \left( \frac{d^n\alpha^A(\omega)}{d\omega^n}\eta^{B}(\omega)+A\leftrightarrow B\right) d\omega \, .
    \label{lambdanfourier}
\end{eqnarray}
To obtain Eq.~~(\ref{Disot}), we have used  that 
$\eta^B(\omega)$ is real-valued and that
$\partial_jG_{ik}=\partial_{i}G_{kj}$. The latter equality holds as the tensor $G_{ij}(\boldsymbol{r})$ defined in Eq.(\ref{E dip}) is symmetric and yields a dipole-induced curl-free electric field $\boldsymbol{E}^{\rm dip}(\boldsymbol{r})$.

At each order, the force contribution $\boldsymbol{F}^{(n)}(\boldsymbol{r}(t)) $ is a product of two factors with distinct physical meanings: a quantum-correlation factor $\Lambda^{(n)}$ accounting for all the internal degrees-of-freedom~(dofs), and a dynamical vector $ \boldsymbol{D}^{(n)}(\boldsymbol{r}(t))$ capturing the influence of the external motion. This structure is the key point underlying the general results presented hereafter.

 
 Memory and dispersion effects become increasingly relevant in $\boldsymbol{F}^{(n)}(\boldsymbol{r}(t))$ as $n$ increases.
For an atom at rest, Eq.~(\ref{Disot}) implies $D_i^{(n)}(\boldsymbol{r})=0$ for $n > 0$ and one recovers the standard London force~\cite{SM}. The same result holds for the particles with an instantaneous (delta-correlated) response function $\alpha(\tau)$. 

For even orders, $\boldsymbol{D}^{(2m)}(\boldsymbol{r})$ is invariant under time reversal $t\to -t$. As quantum friction is by essence an irreversible process, it can only be produced by odd-order dynamical terms $\boldsymbol{D}^{(2m+1)}(\boldsymbol{r})$. The corresponding contributions $\boldsymbol{F}^{(2m+1)}(\boldsymbol{r}) = \boldsymbol{D}^{(2m+1)}(\boldsymbol{r}) \Lambda^{2m+1}$ are thus the only possible candidates for quantum friction effects.

Remarkably, $\Lambda^{2m+1}$ survives only in the presence of a dissipative mechanism. This can be shown with parity considerations. The symmetric correlation function $\eta(\tau)$ is real-valued and even, so that its Fourier transform $\tilde{\eta}(\omega)$ is real-valued and even as well. By the same token, the real (imaginary) part of the susceptibility $\tilde{\alpha}(\omega)$ in the frequency domain is an even (odd) function of the frequency. An inspection of Eq.~(\ref{lambdanfourier}) then shows that only $\mbox{Im}[\tilde{\alpha}(\omega)]$ has the required parity to contribute to $\Lambda^{2m+1}$. The appearance of odd-order quantum force terms
$\boldsymbol{F}^{(2m+1)}(\boldsymbol{r})$ thus requires a susceptibility function with a non-zero imaginary part in the frequency domain.

We have proved on general grounds that the relative motion of dissipationless quantum systems only yields reversible even-order quantum force contributions. Our microscopic theory is thus consistent with the fact that quantum friction requires dissipation at the level of the system's internal dofs.

\textit{Linear dynamical corrections---} The first dynamical correction to the van der Waals force is derived by setting $n=1$ in Eqs.~(\ref{Disot}) and (\ref{lambdanfourier}).  To leading order we ignore dynamical corrections in the response functions, which thus follow the fluctuation-dissipation theorem (FDT) $\eta^{\xi}_T(\omega)=2\coth{\left(\frac{\hbar\omega}{2kT}\right)}\alpha^{\xi}_I{(\omega)}$
where $\alpha^{\xi}_I(\omega)=\mbox{Im}\,\alpha^{\xi}(\omega)$ and $k$ is the Boltzmann constant. Substituting this into Eq.~(\ref{lambdanfourier}) with $n=1$ and integrating by parts we find
\begin{equation}
    \Lambda^{(1)} = -\frac{1}{\pi} \int_{0}^{\infty} d\omega \alpha_I^{A}(\omega)\alpha_I^{B}(\omega)\frac{d}{d\omega}\coth\left(\frac{\hbar\omega}{2kT}\right) \, , \label{Lambda1}
\end{equation}
where we have used 
that $\alpha^{\xi}_I(\omega)$ is odd with  $\alpha_I^{\xi}(\infty)=0.$ In addition, if $\alpha^{\xi}_I(\omega)$ accounts for internal dissipation it must be positive for $\omega\geq 0$ (passivity). As $\coth \left(\frac{\hbar\omega}{2kT}\right)$ is a decreasing function of $\omega$, one finds $\Lambda^{(1)}\geq 0$. 

$\alpha^{\xi}_I(\omega)$, related to the absorption/emission of light, is negligible except near atomic resonances. Let us consider first the high-temperature limit where the dominant contribution comes from atomic transition frequencies $\omega_0$ such that  $\coth (\hbar\omega_0 /2kT)\approx 2kT/(\hbar\omega_0)$ in the FDT relation relating $\eta^{\xi}_T(\omega)$ and $\alpha^{\xi}_T(\omega)$. Eqs.(\ref{serieforca},\ref{Disot},\ref{lambdanfourier}) then yield a force proportional to $kT$ and independent of $\hbar$ --- thus a thermal force unrelated to quantum friction.


Conversely, in the zero-temperature limit, we find $\Lambda^{(1)} \propto \alpha_I^{A}(0)\alpha_I^{B}(0)=0$. Therefore, the first-order dynamical correction only occurs at finite temperature.  For low temperatures ($kT\ll\hbar\omega_0$),  $\partial_{\omega} [ \coth (\hbar\omega/2kT)]$ is significant only for $|\omega|\ll\omega_0$ enabling a low-frequency expansion of $\alpha^{\xi}_I(\omega)$ in Eq.~(\ref{Lambda1}):
\begin{equation}
     \Lambda^{(1)}_{\textup{low T}} \simeq  \frac{2\pi k^2T^2}{3\hbar^2}\left.\frac{d}{d\omega}\alpha_I^{A}\right|_{\omega=0}\left.\frac{d}{d\omega}\alpha_I^{B}\right|_{\omega=0} \, , \label{lambda1lowT}
\end{equation}
where we have used $\int_0^{\infty}d\zeta \zeta^2\frac{d}{d\zeta}\coth\zeta=-\pi^2/6$. For atoms with ultra-violet transitions $\omega_0\sim 10^{16}$ Hz, the approximation (\ref{lambda1lowT}) holds as long as  $T\ll 10^5$ K. Then, the friction-force contribution  $\Lambda^{(1)}$ has a strong quantum character at room temperature.

The dynamical factor $D_i^{(1)}$ is obtained by inserting $dG_{kj}/dt=(\boldsymbol{v}\cdot\boldsymbol{\nabla})G_{kj}$ in Eq.~(\ref{Disot})~\cite{SM}:
\begin{equation}\label{f1}
     \boldsymbol{F}^{(1)}=-\frac{9\hbar}{2(4\pi\varepsilon_0)^2}\left[\frac{2\boldsymbol{v}\cdot\boldsymbol{\hat{r}}}{r^8}\boldsymbol{\hat{r}}+\frac{\boldsymbol{v}}{r^8}\right]\Lambda^{(1)} \,. 
\end{equation}
All the information involving the internal atomic dofs is contained in the $\Lambda^{(1)}$ factor.  The internal atomic structure thus yields a simple scaling factor, without affecting the dependence of  $\boldsymbol{F}^{(1)}$ on atomic velocity and position. Since $\Lambda^{(1)}\geq 0$, we conclude that $\boldsymbol{F}^{(1)}\cdot\boldsymbol{v}\leq 0$ as expected for a dissipative force. 

\textit{Cubic dynamical corrections---} In the zero-temperature limit $\Lambda^{(1)}=0$, turning $\Lambda^{(3)}$ into the leading-order contribution. The latter is obtained  along similar lines as Eqs.~(\ref{Lambda1},\ref{lambda1lowT}):
\begin{equation}
    \Lambda^{(3)}_{\textup{low T}}=-\frac{1}{\pi}\left(\frac{d}{d\omega}\alpha_I^{A}\right)_{\omega=0}\left(\frac{d}{d\omega}\alpha_I^{B}\right)_{\omega=0} \, . \label{lambda3geral}
\end{equation}
From Eqs. (\ref{lambda1lowT},\ref{lambda3geral}), the ratio
\begin{equation}
    \frac{\Lambda^{(3)}_{\textup{low T}}}{\Lambda^{(1)}_{\textup{low T}}} =- \frac{3\hbar^2}{2(\pi kT)^2} \label{universallambda13}
\end{equation}
is a universal function, with $\Lambda^{(3)}_{\textup{low T}} < 0$.  Thus, at low temperatures the ratio $F_i^{(3)}/F^{(1)}_j=D_i^{(3)}\Lambda^{(3)}/(D_j^{(1)}\Lambda^{(1)})$  depends only on the kinematic parameters characterizing the atomic motion and not on internal atomic parameters. To obtain $\boldsymbol{F}^{(3)}$ from Eq. (\ref{Disot}), we use $\dddot{G}_{kj}(\boldsymbol{r}(t)) =  [(\boldsymbol{v}(t)\cdot\boldsymbol{\nabla})^3+3(\boldsymbol{v}(t)\cdot\boldsymbol{\nabla})(\boldsymbol{a}(t)\cdot\boldsymbol{\nabla})
  +(\boldsymbol{j}(t)\cdot\boldsymbol{\nabla})]G_{kj}(\boldsymbol{r})|_{\boldsymbol{r}=\boldsymbol{r}(t)},$
 with $\boldsymbol{a}=\boldsymbol{\ddot{r}}$ representing the relative acceleration and $\boldsymbol{j}=\boldsymbol{\dddot{r}}$ the relative jerk of $B$ with respect to $A$. The ratio between the acceleration contribution and the first one involving only the velocity is roughly $ar/v^2 \sim U_{\rm London}/kT \ll 1,$ when estimating the velocity from the 
 thermal motion and the acceleration from the static London interaction between $A$ and $B$. By a similar argument, the jerk scales as $|\boldsymbol{j}|\sim av/r$ and  yields a contribution of the same order as the acceleration.  Both contributions can be neglected in typical situations. The dominant contribution to $\boldsymbol{F}^{(3)}$ thus stems from the velocity~\cite{SM}:
\begin{eqnarray}
     \boldsymbol{F}^{(3)}&=&-\frac{5\hbar}{2(4\pi\varepsilon_0)^2}\Bigg[\frac{5(\boldsymbol{v}\cdot\boldsymbol{\hat{r}})^2-v^2}{r^{10}}\boldsymbol{v}+\cr\cr
     &+& \frac{85(\boldsymbol{v}\cdot\boldsymbol{\hat{r}})^3-53(\boldsymbol{v}\cdot\boldsymbol{\hat{r}})v^2}{r^{10}}\boldsymbol{\hat{r}}\Bigg]\Lambda^{(3)} \label{f3} \, .
\end{eqnarray}
From Eqs. (\ref{f1},\ref{universallambda13},\ref{f3}),  $|\boldsymbol{F}^{(3)}|/|\boldsymbol{F}^{(1)}|\sim (\hbar\omega_{\rm mot}/kT)^2$ where $\omega_{\rm mot}=v/r$ is a characteristic frequency determined by the motion and the interatomic distance. For $v\sim 10^{3}$ m/s and $r\sim 1$ nm we obtain $|\boldsymbol{F}^{(3)}|/|\boldsymbol{F}^{(1)}|\sim 10^{-3}$  for room temperature. Hence, for realistic scenarios one has $\hbar\omega_{\rm mot}\ll kT\ll \hbar\omega_0$. The first inequality implies that the dominant friction is approximately linear in the velocity. The inequality $ kT\ll \hbar\omega_0$ ensures that this friction has a quantum character (see Eqs. (\ref{lambda1lowT},\ref{f1})). Remarkably, even though the excited state population is exponentially suppressed by a  Boltzmann factor $e^{- k T/\hbar \omega_0}$, it still induces the dominant contribution to the quantum friction force for  $kT\gg\hbar\omega_{\rm mot}$. 

\begin{figure}[h!]
    \centering
    \includegraphics[width=1\columnwidth]{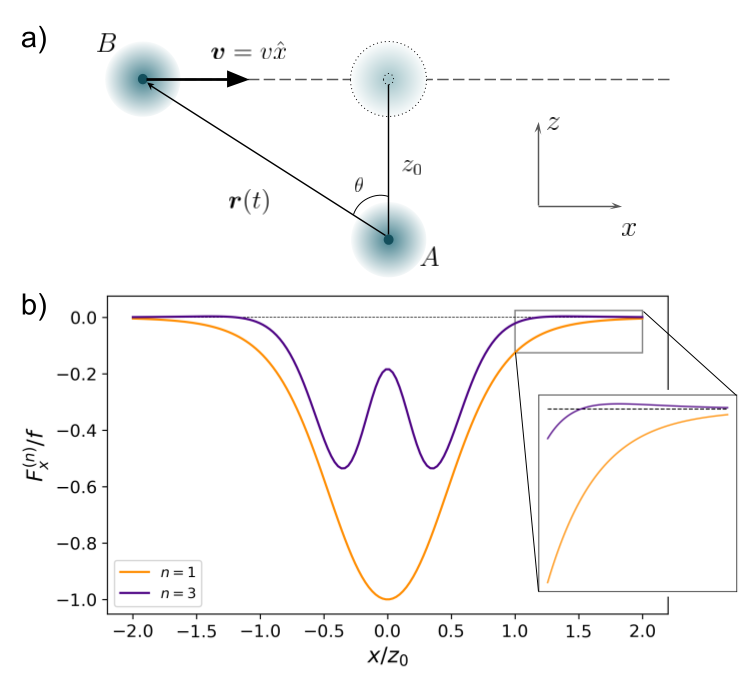}
    \caption{a) Schematics of atom $B$ moving in uniform motion with respect to atom $A$. b) Plot of $F^{(1)}_x$ and $F^{(3)}_x$ versus the relative lateral position $x$ normalized by the minimum distance $z_0$. The forces are normalized by $f=|F^{(1)}(\boldsymbol{r}=z_0\boldsymbol{\hat{z}})|$ and we chose parameters of temperature, velocity and distance such as $(kTz_0/\hbar v)^2=0.05$. The inset reveals that  $F^{(3)}_x$ can be locally positive, corresponding to energy transfer to the CM.}
    \label{fig:figure}
\end{figure}

Fig. (~\ref{fig:figure}) pictures the quantum friction contributions ${F}^{(1)}_x$ and ${F}^{(3)}_x$ 
along the trajectory $(x=vt,z_0).$
We consider the regime $kT \ll \hbar \omega_{\rm mot}$ to enhance the third order contribution. 
To estimate the frequency $\omega_{\text{mot}} = v/r$, we take $r \simeq z_0$ with $z_0$ denoting the impact parameter. Defining $\theta$ as the angle between $\boldsymbol{r}$ and the z axis, we see from Eq.(\ref{f3}) that $\boldsymbol{F}^{(3)}$ displays an unexpected gain ($\boldsymbol{F}^{(3)}\cdot\boldsymbol{v}>0$) for $\theta>49.88^o$, while $\boldsymbol{F}^{(1)}$ is strictly dissipative along the trajectory. This angle is independent of $v$, $z_0$ or the atoms involved.
Below, we derive general theorems showing that such local gain does not contradict quantum friction.

\textit{Total work performed on the moving atom.} Let us consider a scattering process, satisfying the condition $|\boldsymbol{r}(t\to\pm\infty)|\to \infty$, and denote $W^{(n)}$ the total work performed by the force term $\boldsymbol{F}^{(n)}$. One expects  $W_{\rm tot} = \sum_n W^{(n)} \leq 0$. Otherwise, the net effect would be a conversion of internal energy into CM kinetic energy -- in contradiction with  the second law of thermodynamics. We now demonstrate general theorems on $W^{(n)}$ without the need of evaluating explicitly $D_i^{(n)}$ or specifying any internal atomic model. We analyze separately the cases $n=2k$ (even) and $n=2k+1$ (odd) with $k \in \mathbb{N}.$

\textbf{Theorem 1} In an arbitrary scattering process the total work performed by $\boldsymbol{F}^{(2k)}$ is zero.

\textbf{Theorem 2} In an arbitrary scattering process the total work performed by $\boldsymbol{F}^{(2k+1)}$  has the sign of $(-1)^{k+1}\Lambda^{(2k+1)}$.

\textit{Demonstration:}  The total work performed by $\boldsymbol{F}^{(n)}$ is given by $\Lambda^{(n)}\int_{-\infty}^{\infty}\boldsymbol{D}^{(n)}\cdot\boldsymbol{v}dt$, with the dynamical factor given in Eq.~(\ref{Disot}). Using  $(\boldsymbol{v}\cdot\boldsymbol{\nabla})G_{lm}=dG_{lm}/dt$ we obtain
\begin{equation}
    W^{(n)} =\frac{\hbar\Lambda^{(n)}(-1)^n}{2n!}\!\!\int_{-\infty}^{\infty} \!\! dt\left(\frac{d}{dt}G_{lm}\right)\left(\frac{d^n}{dt^n}G_{lm}\right) \, . \label{totalwork}
 \end{equation}
When integrating by parts we can discard the surface terms due to the scattering condition.
For $n=2k$, we take $2k-1$ integrations by parts in Eq.~(\ref{totalwork})  and find $W^{(2k)}=-W^{(2k)}$ demonstrating Theorem 1. For $n=2k+1,$ we make $k$ integrations by parts:
\begin{equation*}
    W^{(2k+1)} =\frac{(-1)^{k+1}\Lambda^{(2k+1)}}{2(2k+1)!}\!\!\int_{-\infty}^{\infty} \!\!dt\sum_{l,m}\left(\frac{d^{k+1}}{dt^{k+1}}G_{lm}\right)^2 \! , 
 \end{equation*}
thus proving Theorem 2.

The fact that $W^{(2k)}=0$ is in agreement with the absence of dissipative character for $\boldsymbol{F}^{(2k)}$ discussed previously. Although 
$\boldsymbol{F}^{(2k)}$ can  locally inject or extract energy, 
there is no overall energy transfer when integrating over the entire trajectory. 
All the effect of friction manifests itself in the odd terms. We demonstrated that $\Lambda^{(1)} \geq 0$, so that theorem 2 implies $W^{(1)}\leq 0$, as expected for a dissipative force. For $n=1$ the integrand in Eq.~(\ref{totalwork}) is positive, enabling a stronger statement: the power is negative at all times along the trajectory. 
In the low temperature limit, the cubic term yields the dominant contribution. We have shown that $\Lambda^{(3)} < 0$ and thus $W^{(3)} < 0$. Therefore, although $\boldsymbol{F}^{(3)}$ can briefly do a positive work, 
its overall effect is to dissipate the energy of the CM.

One expects that $\Lambda^{(2k+1)} = (-1)^{k}|\Lambda^{(2k+1)}|$ also holds at higher orders ($ k \geq 2$) in order to yield $W^{(2k+1)}\leq 0$. This can be readily verified in the zero-temperature limit assuming the FDT relations. We model the polarizability of atom $A$ as a multilevel Lorentz model with resonance
frequencies $\omega_{a}$ and linewidths $\gamma_{a}.$ Using the index $b$ to denote the transitions of atom $B,$ we obtain~\cite{SM}
\begin{equation}
    \Lambda^{(2k+1)} =(-1)^k \sum_{a,b}\frac{\alpha_a(0)\alpha_b(0)\gamma_a\gamma_b}{4\pi}\sum_{j=0}^{k-1} \frac{(2k-2j)!(2j+2)!}{\omega_a^{(2k-2j)}\omega_b^{(2j+2)}} \, , \label{lambda2k+1}
\end{equation}
where the positive parameter $\alpha_a(0)$ represents the contribution of transition $a$ to the static polarizability when  $\gamma_a\ll \omega_a.$

\textit{Connection with macroscopic quantum friction.} Let us consider that atom $B$ is moving with uniform velocity $\boldsymbol{v}$ keeping a constant distance $z_0$ from a rarefied medium filling the region $z<0$. The medium contains $N$ atoms $A$ per unit volume. The force $\boldsymbol{F}_{\rm net}$ exerted by the medium on atom $B$ is then given by the pairwise integration of the force obtained for two atoms ~\cite{FootnoteRetardation}. 

Once again, contributions arising from $D_i^{(n)}$ behave very differently for even and odd $n$. From general symmetry considerations, it can be shown that the average of even (odd) $n$ yields a force perpendicular (parallel) to the motion~\cite{SM}. Indeed, our microscopic model also captures  the main features of this macroscopic situation: all contributions to the  macroscopic friction force arise from microscopic terms with odd $n$.  The explicit calculation for $n=1,3$ yields
\begin{eqnarray}
    \boldsymbol{F}_{\rm net}^{(1)}&=&-\frac{3N\hbar\Lambda^{(1)}}{80\pi\varepsilon_0^2}\frac{\boldsymbol{v}}{z_0^5}; \: \: 
    \boldsymbol{F}_{\rm net}^{(3)} = \frac{21N\pi\hbar\Lambda^{(3)}v^2}{64(4\pi\varepsilon_0)^2z_0^7}\boldsymbol{v} \, . \label{f3plate}
\end{eqnarray}
Since $\Lambda^{(1)}\geq 0$ and $\Lambda^{(3)}<0$, both components have a dissipative character. 
At a given time, the power transferred through $\boldsymbol{F}^{(3)}$  by a fraction of the atoms of the medium is actually 
positive. However, the net effect of the entire medium as captured by 
 $\boldsymbol{F}_{\rm net}^{(3)}$ remains dissipative at all times.


\textit{Final remarks---} We have provided a self-consistent theory for quantum friction at the atomic level. 
In the short-distance limit, where this effect is most relevant, quantum friction arises mainly from the delay in the atomic response. As expected from its irreversible character, the microscopic quantum friction force only survives in the presence of an internal dissipative mechanism. Our perturbative expansion of the microscopic quantum friction explicitly separates the roles of the internal atomic dofs from the CM motion. The former are gathered in quantum correlation factors $\Lambda^{(n)}$, whose sign depends on the parity of $n$ and can be determined without any specific model for the internal atomic structure. We have shown that odd-order contributions are the only irreversible terms, thus fully capturing microscopic quantum friction.

At zero temperature the main contribution $\boldsymbol{F}^{(3)}$ to microscopic quantum friction has a cubic dependence towards velocity, while at finite temperatures the leading term $\boldsymbol{F}^{(1)}$ is linear in the velocity. Under usual temperature conditions, even $\boldsymbol{F}^{(1)}$  is actually still dominated by quantum, rather then thermal, fluctuations.

The microscopic quantum friction force presents counterintuitive features, such as the possibility for higher-order contributions to provide positive work over a fraction of the CM trajectory.  Nevertheless, we have shown on general grounds that the dissipative character of quantum friction terms prevails when considering the work performed over a full scattering process. 

Our microscopic \textit{ab-initio} theory evidenced several universal and material-independent features of quantum friction, allowing us to obtain results that hold in the non-equilibrium regime. It provides a deeper understanding of macroscopic quantum friction at the atomic scale and can help design future strategies 
and experiments to enhance this effect.

{\bf Acknowledgments.} The authors thank
Conselho Nacional de Desenvolvimento
Científico e Tecnológico (CNPq–Brazil), Coordenação
de Aperfeiçamento de Pessoal de Nível Superior
(CAPES–Brazil), and the Research
Foundation of the State of Rio de Janeiro (FAPERJ)
for finantial support.







\bibliography{pubs}

@misc{SM,
  note = "See Supplemental Material at
    [URL will be inserted by publisher] for the calculation of the correlation between the fluctuatig dipoles, for the evaluation of the dynamical vector and the quantum-correlation factor and for the spatial angular average of the force in the connection with the macroscopic quantum friction."}

@article{fact_or_fiction,
doi = {10.1088/1367-2630/12/3/033028},
url = {https://dx.doi.org/10.1088/1367-2630/12/3/033028},
year = {2010},
month = {mar},
publisher = {},
volume = {12},
number = {3},
pages = {033028},
author = {Pendry, J B},
title = {Quantum friction–fact or fiction?},
journal = {New Journal of Physics}
}

@article{no_friction,
doi = {10.1088/1367-2630/11/3/033035},
url = {https://dx.doi.org/10.1088/1367-2630/11/3/033035},
year = {2009},
month = {mar},
publisher = {},
volume = {11},
number = {3},
pages = {033035},
author = {Philbin, T G and Leonhardt, U},
title = {No quantum friction between uniformly moving plates},
journal = {New Journal of Physics}
}

@article{friction_experimental,
  title = {Quantum Friction},
  author = {Volokitin, A. I. and Persson, B. N. J.},
  journal = {Phys. Rev. Lett.},
  volume = {106},
  issue = {9},
  pages = {094502},
  numpages = {4},
  year = {2011},
  month = {Mar},
  publisher = {American Physical Society},
  doi = {10.1103/PhysRevLett.106.094502},
  url = {https://link.aps.org/doi/10.1103/PhysRevLett.106.094502}
}

@book{IsraelachviliBook,
    author = {Jacob N. Israelachvili},
    title = {Intermolecular and Surface Forces},
    publisher = {Academic Press},
    year = {2011}
}

@article{Buhmann09,
  title = {Casimir-{P}older forces on moving atoms},
  author = {Scheel, Stefan and Buhmann, Stefan Yoshi},
  journal = {Phys. Rev. A},
  volume = {80},
  issue = {4},
  pages = {042902},
  numpages = {11},
  year = {2009},
  month = {Oct},
  publisher = {American Physical Society},
  doi = {10.1103/PhysRevA.80.042902},
  url = {https://link.aps.org/doi/10.1103/PhysRevA.80.042902}
}

@article{Sanders2019,
    author = {Stephen Sanders and Wilton J. M. Kort-Kamp and Diego A. R. Dalvit and Alejandro Manjavacas},
    title = {Nanoscale transfer of angular momentum mediated by the Casimir torque},
    journal = {Commun.Phys.},
    volume = {2},
    pages = {71},
    year = {2019}
}

@article{Amaral2025,
    author = {H. S. G. Amaral and P. P. Abrantes and F. Impens and P. A. Maia Neto and R. de Melo e Souza},
    title = {Tailoring the van der Waals interaction with rotation},
    journal = {Phys.Rev.Lett.},
    volume = {135},
    pages = {243601},
    year = {2025}
}

@article{Manjavacas2010,
    author = {Manjavacas, A. and García de Abajo, F. J.},
    title = {Vacuum Friction in Rotating Particles},
    journal = {Phys. Rev. Lett.},
    volume = {105},
    pages = {113601},
    year = {2010},
    doi = {10.1103/PhysRevLett.105.113601},
    url = {https://journals.aps.org/prl/abstract/10.1103/PhysRevLett.105.113601},
}

@article{Xu2021,
    author = {Xu, Zhujing and Jacob, Zubin and Li, Tongcang},
    title = {Enhancement of rotational vacuum friction by surface photon tunneling},
    journal = {Nanophotonics},
    volume = {10},
    pages = {537},
    year = {2021},
    doi = {https://doi.org/10.1515/nanoph-2020-0391},
    url = {https://www.degruyter.com/document/doi/10.1515/nanoph-2020-0391/html},
}

@article{Reiche2020,
    author = {D. Reiche and F. Intravaia and J.-T. Hsiang and K. Busch and B.-
L. Hu},
    title = {Nonequilibrium thermodynamics of quantum friction},
    journal = {Phys. Rev. A},
    volume = {102},
    pages = {050203(R)},
    year = {2020}
}

@article{Milton2016,
    author = {K. A. Milton and J. S. Høye and I. Brevik},
    title = {The reality of Casimir
friction},
    journal = {Symmetry},
    volume = {8},
    pages = {29},
    year = {2016}
}

@article{Reiche2017,
    author = {D. Reiche and  D. A. R. Dalvit and K. Busch and F. Intravaia},
    title = {Spatial dispersion in atom-surface quantum friction},
    journal = {Phys. Rev. B},
    volume = {95},
    pages = {155448},
    year = {2017}
}

@article{Farias2018,
    author = {M. Belén Farias and Wilton J. M. Kort-Kamp and Diego A. R. Dalvit},
    title = {Quantum friction in two-dimensional topological materials},
    journal = {Phys. Rev. B},
    volume = {97},
    pages = {161407},
    year = {2018}
}

@article{Guo2021,
    author = {X. Guo and K. A. Milton and G. Kennedy and W. P. McNulty and N.
Pourtolami and Y. Li},
    title = {Energetics of quantum vacuum friction:
Field fluctuations},
    journal = {Phys. Rev. D},
    volume = {104},
    pages = {116006},
    year = {2021}
}

@article{Guo2022,
    author = {X. Guo and K. A. Milton and G. Kennedy and W. P. McNulty and N.
Pourtolami and Y. Li},
    title = {Energetics of quantum vacuum friction:
Field fluctuations},
    journal = {Phys. Rev. D},
    volume = {106},
    pages = {016008},
    year = {2022}
}

@article{Lombardo2017,
    author = {F. C. Lombardo and P. I. Villar},
    title = {Geometric phase corrections on a moving particle in front of a dielectric mirror},
    journal = {EPL},
    volume = {118},
    pages = {50003},
    year = {2017}
}

@article{Lombardo2021,
    author = {F. C. Lombardo and R. S. Decca and L. Viotti and P. I. Villar},
    title = {Detectable Signature of Quantum Friction on a Sliding Particle in Vacuum},
    journal = {Adv. Quant. Tech.},
    volume = {4},
    pages = {2000155},
    year = {2021}
}

@article{Viotti2019,
    author = {L. Viotti and M. B. Farias and P. I. Villar and F. C. Lombardo},
    title = {Thermal corrections to quantum friction and decoherence: A closed-time-path approach to atom-surface interaction},
    journal = {Phys.Rev.D},
    volume = {99},
    pages = {105005},
    year = {2019}
}

@article{Xu2024,
    author = {Zhujing Xu and Peng Ju and Kunhong Shen and Yuanbin Jin and Zubin Jacob and Tongcang Li},
    title = {Observation of non-contact Casimir friction},
    journal = {arXiv:2403.06051},
    year = {2024}
}

@article{Farias2020,
    author = {M. B. Farias and F. C. Lombardo and A. Soba and P. I. Villar and R. S.
Decca},
    title = {Towards detecting traces of non-contact quantum friction in the corrections of the accumulated geometric phase},
    journal = {npj Quantum Inf.},
    volume = {6},
    pages = {25},
    year = {2020}
}

@article{Reiche2022,
    author = {D. Reiche and F. Intravaia and K. Busch},
    title = {Wading through the void: Exploring quantum friction and nonequilibrium fluctuations},
    journal = {APL Photonics},
    volume = {7},
    pages = {030902},
    year = {2022}
}

@article{Oue2025,
    author = {Daigo Oue and Boris Shapiro and Mário G. Silveirinha},
    title = {Quantum friction near the instability threshold},
    journal = {Phys.Rev. B},
    volume = {111},
    pages = {075403},
    year = {2025}
}

@article{Oue2024,
    author = {Daigo Oue and J. B. Pendry and Mário G. Silveirinha},
    title = {Stable-to-unstable transition in quantum friction},
    journal = {Phys.Rev.Research},
    volume = {6},
    pages = {043074},
    year = {2024}
}

@article{Shaukat2025,
    author = {Muzzamal I. Shaukat and Mário G. Silveirinha},
    title = {Impact of chiral transitions in quantum friction},
    journal = {Phys.Rev.A},
    volume = {111},
    pages = {032202},
    year = {2022}
}

@article{Gao2025,
    author = {Xinchen Gao and Zhenbin Gong and Hongli Li and Zhao Liu and Weishan Yan and Qingkai Zheng and Kexin Ren and, Wenchao Wu and Junyan Zhang },
    title = {Pseudo-Landau levels splitting triggers quantum friction at folded graphene edge},
    journal = {Nature Communications},
    volume = {16},
    pages = {5558},
    year = {2025}
}

@article{Farias2019,
    author = {M. B. Farias and C. D. Fosco and F. C. Lombardo and F.D. Mazzitelli},
    title = {Motion induced radiation and quantum friction for a moving atom},
    journal = {Phys.Rev.D},
    volume = {100},
    pages = {036013},
    year = {2019}
}

@article{Franca2025,
    author = {O. J. Franca and Fabian Spallek and Steffen M. Giesen and Robert Berger and Kilian Singer and Stefan Aull and Stefan Yoshi Buhmann},
    title = {Spectroscopic footprints of quantum friction in nonreciprocal and chiral media},
    journal = {Phys.Rev.A},
    volume = {112},
    pages = {012803},
    year = {2025}
}

@article{Fernandez2024,
    author = {Aitor Fernández and César D. Fosco},
    title = {Quantum friction for a scalar model: Spatial dependence and higher orders},
    journal = {Annals of Physics},
    volume = {463},
    pages = {169635},
    year = {2024}
}

@article{Kavokine2022,
    author = {Nikita Kavokine and Marie-Laure Bocquet and Lydéric Bocquet},
    title = {Fluctuation-induced quantum friction in nanoscale water flows},
    journal = {Nature},
    volume = {602},
    pages = {84},
    year = {2022}
}

@article{Ho2025,
    author = {Jeongwon Ho and O-Kab Kwon and Sang-Heon Yi},
    title = {Quantum inhomogeneous field theory: Unruh-like effects and bubble wall friction},
    journal = {JHEP},
    volume = {2025},
    pages = {58},
    year = {2025}
}

@article{Fosco2021,
    author = {C. D. Fosco and F. C. Lombardo and F.D. Mazzitelli},
    title = {Motion-Induced Radiation Due to an Atom in the Presence of a Graphene Plane},
    journal = {Universe},
    volume = {7},
    pages = {158},
    year = {2021}
}

@article{Donaire2016,
    author = {M. Donaire and A. Lambrecht},
    title = {Velocity-dependent dipole forces on an excited atom},
    journal = {Phys.Rev.A},
    volume = {93},
    pages = {022701},
    year = {2016}
}

@article{Impens13a,
    author = {Impens, F. and Behunin, R. O. and Ccapa-Ttira, C.  and Maia Neto, P. A.},
    title = {Non-local double-path Casimir phase in atom interferometers},
    journal = {EPL},
    volume = {101},
    pages = {60006},
    year = {2013},
    doi = {10.1209/0295-5075/101/60006},
    url = {https://iopscience.iop.org/article/10.1209/0295-5075/101/60006/meta},
}

@article{Matos2020,
  title = {Quantum Vacuum Sagnac Effect},
  author = {Matos, G.C. and de Melo e Souza, R. and Neto, P.A. Maia and Impens, F.},
  journal = {Phys. Rev. Lett.},
  volume = {127},
  issue = {27},
  pages = {270401},
  numpages = {6},
  year = {2021},
  month = {Dec},
  publisher = {American Physical Society},
  doi = {10.1103/PhysRevLett.127.270401},
  url = {https://link.aps.org/doi/10.1103/PhysRevLett.127.270401}
}

@misc{FootnoteRetardation,
  note = "Although this integration involves pair of atoms at arbitrary distances, for $\omega z_0/c\ll 1$ the main contribution comes from a spatial region where the non-retarded regime applies."}

@article{MILTON2025130475,
title = {Perspectives on quantum friction, self-propulsion, and self-torque},
journal = {Physics Letters A},
volume = {545},
pages = {130475},
year = {2025},
issn = {0375-9601},
doi = {https://doi.org/10.1016/j.physleta.2025.130475},
url = {https://www.sciencedirect.com/science/article/pii/S0375960125002567},
author = {Kimball A. Milton and Nima Pourtolami and Gerard Kennedy}
}
\end{document}


\title{Supplemental Material: Microscopic Quantum Friction}

\author{Pedro H. Pereira}
    \affiliation{Instituto de Física, Universidade Federal Fluminense, 24210-346, RJ, Brazil}
    \email{pedro_h@id.uff.br}
\author{F. Impens}
\author{C. Farina}
\author{P. A. Maia Neto}
\affiliation{Instituto de F\'{i}sica, Universidade Federal do Rio de Janeiro, 21941-972, Rio de Janeiro, Rio de Janeiro,  Brazil}

\author{R. de Melo e Souza}
\email{reinaldos@id.uff.br}
    \affiliation{Instituto de Física, Universidade Federal Fluminense, 24210-346, RJ, Brazil}

\begin{abstract}
This Supplemental Material provides additional details on (I) the non-retarded regime of quantum friction, (II) the derivation of the correlation of the dipole operators, (III) a covariant calculation for the orders $n=1,2,3$ of the dynamical vector, (IV) the discussion on the sign of the quantum-correlation factor, and (V) the evaluation of the angular average of the force in connection with the macroscopic quantum friction. 
\end{abstract}

\maketitle

\section{Non-retarded quantum friction}  

We justify here with a simple argument the non-retarded approximation to evaluate microscopic quantum friction.  Let us consider two interacting atoms with a relative instantaneous velocity $v(t)$ and distance $r(t)$. At a fundamental level, the dispersion interaction is mediated by the exchange of virtual photons -- a three-step process involving two free-space propagations and a scattering over one of the particles. These stages follow distinct time scales, namely a retardation time $\tau_{\rm ret} \sim 2 r/c$ for the back-and-forth light travel, and a dispersive response time $\tau_{\rm dis} \sim 1/ \nu_0$  corresponding to the particle response  ($\nu_0$ is the frequency of the considered atomic transition). The corresponding Casimir force corrections $\epsilon =\Delta F/F $ scale as the relative distance change $\Delta r/r$ during each stage (either light propagation or particle response) $\epsilon \sim v \tau /r$.  Thus, the dynamical corrections typically involve two distinct small parameters, $\epsilon_{\rm ret} = 2 v/c$ and $\epsilon_{\rm dis} = v/(r \nu_0),$ respectively associated to retardation and dispersive effects. As $\epsilon_{\rm ret}/ \epsilon_{\rm dis} = 2 r/\lambda_0$ with $\lambda_0$ the transition wavelength, our analysis shows that dispersive corrections predominate at short distances $r \ll \frac 1 2 \lambda_0$. Considering two atoms with low-frequency (for instance micro-wave) transitions, one sees that retardation effects are insignificant when investigating vdW force corrections at the nanoscale ($r \sim 10-100 {\rm nm}$).

\section{Correlation between dipole fluctuations}

    We evaluate here the leading-order contribution to the crossed dipole correlation function  $\langle d_m^B(t)d_l^A(t)\rangle$. The time evolution of the dipole operators can be evaluated perturbatively~\cite{Santos24}
\begin{equation}
     d_m^B(t) = d_m^{(0)B}(t)+\frac{i}{\hbar}\int_{-\infty}^tdt'[d_m^{(0)B}(t),d_j^{(0)B}(t')]G_{jk}(\boldsymbol{r}(t'))d_k^{(0)A}(t') \, , \label{evoldip}
\end{equation}
with an implicit summation on repeated indices. $\boldsymbol{d}^{(0) \xi}(t)$ denotes the free-evolved  dipole operator of the atom $\xi=A,B$ in the interaction picture. Eq.~(\ref{evoldip}) means that to first (linear) order of perturbation theory, the induced dipole  $\boldsymbol{d}^B(t)$ arises from the electric field $G_{jk}(\boldsymbol{r}(t'))d_k^{(0)A}(t')$ generated by the dipole $A$ at the position of $B$ during the time interval $t'\in[-\infty,t]$. An analogous expression holds for $\boldsymbol{d}^A(t)$, so that the crossed dipole correlation function reads:

%
\begin{equation}
    \langle d_m^B(t)d_l^A(t)\rangle  =\int_0^tdt' \Bigg(\alpha_{mj}^B(t-t')G_{jk}(\boldsymbol{r}(t'))\langle d_k^{(0)A}(t')d_l^{(0)A}(t)\rangle+
    \langle d_m^{(0)B}(t)d_j^{(0)B}(t')\rangle G_{jk}(\boldsymbol{r}(t'))\alpha_{lk}^A(t-t')\Bigg) \, .
\end{equation}
where we used that the atoms have no permanent dipole moment $\langle d_m^{(0) B}(t)d_l^{(0) A}(t)\rangle = 0$, as well as  the symmetry of the tensor $G_{jk}$.  The polarizability tensor is defined by
\begin{equation}\label{alpha}
    \alpha_{mj}^\mathcal{\xi}(t-t')=\frac{i}{\hbar}\theta(t-t')\langle[d^{(0)\xi}_m(t),d^{(0)\xi}_j(t')]\rangle\,; \,\,\, (\xi=A,B) \, .
\end{equation}
We also define the fluctuation tensor (also denominated Hadamard Green function) by
\begin{equation}
    \eta_{kl}^\mathcal{\xi}(t-t')=\frac{1}{\hbar}\langle\{d^{(0)\xi}_k(t),d^{(0)\xi}_l(t')\}\rangle \,; \,\,\, (\xi=A,B) \, .
\end{equation}
Using the relations $\langle d_k^{(0)A}(t')d_l^{(0)A}(t)\rangle = \frac{\hbar }{2}\left(\eta_{lk}^A(t-t')+i\chi_{lk}^A(t-t')\right)$ and $\langle d_m^{(0)B}(t)d_j^{(0)B}(t')\rangle = \frac{\hbar }{2}\left(\eta_{mj}^B(t-t')-i\chi_{mj}^B(t-t')\right)$ for $t'<t$, we obtain
\begin{equation}
 \langle d_m^B(t)d_l^A(t)\rangle = \frac{\hbar}{2} \int_0^t dt' G_{kj}(\boldsymbol{r} (t'))\Big(\chi_{mj}^B(t-t')\eta^A_{lk}(t-t')+\eta^B_{mj}(t-t')\chi_{lk}^A(t-t')\Big) \, .
\end{equation}

\section{Covariant derivation of the microscopic quantum friction force} 


Below we use symmetry arguments to evaluate tensorial quantities relevant for microscopic quantum friction. Let us define:
\begin{equation}\label{defGcal}
    \mathcal{G}_{ij_1\cdots j_n}(\boldsymbol{r})\equiv (\partial_{i}G_{lm}(\boldsymbol{r}))\partial_{j_1}\cdots\partial_{j_n}G_{lm}(\boldsymbol{r}) \, ,
\end{equation}
where
\begin{equation}
    G_{lm}(\boldsymbol{r}) =  \frac{3\hat{r}_l\hat{r}_m-\delta_{lm}}{4\pi\varepsilon_0r^3} \, . \label{E dipSM}
\end{equation}
%
The static van-der-Waals force is expressed in terms of this tensor as 

\begin{equation}\label{F0}
F_i^{(0)}(\boldsymbol{r}) = \frac{\hbar}{2}\mathcal{G}_{i}(\boldsymbol{r})\Lambda^{(0)}
\end{equation}

Similarly, the contributions to microscopic quantum friction of order $n=1,3$ (see main text) can be readily expressed as:
%
\begin{eqnarray}\label{d0}
    F_i^{(1)}(\boldsymbol{r})&=&-\frac{\hbar}{2}v_k\mathcal{G}_{ik}(\boldsymbol{r})\Lambda^{(1)} \, , \label{d1cov}\\
    F_i^{(3)}(\boldsymbol{r}) &=&-\frac{\hbar}{12}\Big(v_kv_nv_s\mathcal{G}_{ikns}(\boldsymbol{r})+3v_ka_n\mathcal{G}_{ikn}(\boldsymbol{r})+j_k\mathcal{G}_{ik}(\boldsymbol{r})\Big) \Lambda^{(3)}\, ,\label{d3cov}
\end{eqnarray}
where $\boldsymbol{v},\boldsymbol{a}$ and $\boldsymbol{j}$ denotes the relative velocity, acceleration and jerk, respectively and $\Lambda^{(n)}$ are the quantum-correlation factors defined in the main text.

\vskip 0.3 cm

\textit{Obtaining $\mathcal{G}_i(\boldsymbol{r})$}

\vskip 0.3 cm

From the definition (\ref{defGcal}) and Eq.(\ref{E dipSM}), we have:
\begin{equation}\label{gcali}
    \mathcal{G}_i(\boldsymbol{r})= \frac{(3\hat{r}_l\hat{r}_m-\delta_{lm})}{4\pi\varepsilon_0r^3} \partial_i\frac{3\hat{r}_l\hat{r}_m-\delta_{lm}}{4\pi\varepsilon_0r^3}  \, .
\end{equation}
%
As a first-order tensor, symmetry considerations show that $\mathcal{G}_i(\boldsymbol{r})$ must be of the form
$\mathcal{G}_i(\boldsymbol{r})=f(r) \hat{r}_i $ with $f(r)$ a scalar function depending on the distance $r$ only. Dimensional analysis yields for $f(r)$ the following scaling law:
\begin{equation}\label{gcalicov}
    \mathcal{G}_i(\boldsymbol{r})= \frac{\kappa_0\hat{r}_i}{(4\pi\varepsilon_0)^2r^7} \, ,
\end{equation}
%
$\kappa_0$ is a constant that can be determined by evaluating Eq.(\ref{gcali}) for the particular case  $\boldsymbol{r}=x\boldsymbol{\hat{x}}$. $\mathcal{G}_x(\boldsymbol{r})$ is then the only nonzero Green's function component. In this specific case, $\hat{\boldsymbol{r}}=\hat{\mathbf{x}}$ and $\partial_x \hat{\boldsymbol{r}} =\boldsymbol{0}$, so that both factors $3\hat{r}_l\hat{r}_m-\delta_{lm}$ of Eq.\eqref{gcali} can be included within the derivative $\partial_i[...] \equiv \partial_x[...]$. As $(3\hat{r}_l\hat{r}_m-\delta_{lm})(3\hat{r}_l\hat{r}_m-\delta_{lm})=6,$ one finds:
%
\begin{eqnarray}
    \mathcal{G}_x(x\boldsymbol{\hat{x}}) &=& -\frac{18}{(4\pi\varepsilon_0)^2x^7}\, .   \label{resultsymmetryforce0}
\end{eqnarray}
Comparison of Eqs.~(\ref{gcalicov},\ref{resultsymmetryforce0}) yields $\kappa_0=-18$. 
%
%
Substituting Eq. (\ref{gcalicov}) in Eq. (\ref{F0}) and using Eq. (\ref{lambdanfourier}) of the main text for $\Lambda^{(0)}$, we obtain
\begin{equation}
    \boldsymbol{F}^{(0)}=-\frac{9\hbar\boldsymbol{\hat{r}}}{32\pi^3\varepsilon_0^2r^7}\int _{-\infty}^{\infty}
    \left(\alpha^A(\omega)\eta^{B}(\omega)+A\longleftrightarrow B\right) d\omega \, ,
\end{equation}
This result coincides with the expected dispersion force (also known as London force) between two atoms  in the non-retarded static case~\cite{Santos24}.

\vskip 0.3 cm

\textit{Calculation of $\mathcal{G}_{ik}(\boldsymbol{r})$}

\vskip 0.3 cm

From Eqs.(\ref{defGcal},\ref{E dipSM}), this tensor reads:
\begin{equation}\label{gcalik1}
    \mathcal{G}_{ik}(\boldsymbol{r}) = \left[\partial_k\frac{(3\hat{r}_l\hat{r}_m-\delta_{lm})}{4\pi\varepsilon_0r^3} \right]\partial_i\frac{3\hat{r}_l\hat{r}_m-\delta_{lm}}{4\pi\varepsilon_0r^3}  \, .
\end{equation}
%
 As a second-rank tensor, and by virtue of dimensional considerations, $\mathcal{G}_{ik}(\boldsymbol{r})$ must be of the general form:
\begin{equation}
   \mathcal{G}_{ik}(\boldsymbol{r})= \frac{\kappa_{1}\hat{r}_i\hat{r}_k+\kappa_2\delta_{ik}}{(4\pi\varepsilon_0)^2r^8} \label{symmetryforce11}\, .
\end{equation}
%

To determine the dimensionless constants $\kappa_1$ and $\kappa_2$, we consider two particular configurations: (C1): $i,k=x$ and $\boldsymbol{r}=x\boldsymbol{\hat{x}}$ and (C2) $i,k=x$ and $\boldsymbol{r}=y\boldsymbol{\hat{y}}$. Substituting these choices into Eq.~(\ref{symmetryforce11}) we obtain the system of two equations for $\kappa_1,\kappa_2$:
\begin{eqnarray}
      \mathcal{G}_{xx}(x\boldsymbol{\hat{x}})  &=& \frac{\kappa_1+\kappa_2}{(4\pi\varepsilon_0)^2x^8} \;\; \mbox{(C1)} \, , \label{C1} \\
            \mathcal{G}_{xx}(y\boldsymbol{\hat{y}})&=& \frac{\kappa_2}{(4\pi\varepsilon_0)^2y^8} \;\; \mbox{(C2)} \, . \label{C2}
\end{eqnarray}
which are then determined by evaluating $\mathcal{G}_{xx}(\boldsymbol{\hat{r}})$ in Eq. (\ref{gcalik1}) on the positions associated to $(C1),(C2)$.\\

\textit{Calculation of (C1)}\\


As $\partial_x \hat{\boldsymbol{r}}(x,y,z) |_{x \neq 0, y=z=0} = \boldsymbol{0},$ one can once again include in Eq. (\ref{gcalik1}) the tensor $(3\hat{r}_l\hat{r}_m-\delta_{lm})$ within the derivative $\partial_i$ and use $(3\hat{r}_l\hat{r}_m-\delta_{lm})^2=6$:\\

\begin{equation}
          \mathcal{G}_{xx}(x\boldsymbol{\hat{x}}) =\frac{6}{(4\pi\varepsilon_0)^2}\left(\partial_x\frac{1}{x^3}\right)^2=\frac{54}{(4\pi\varepsilon_0)^2x^8} \, . \label{resultsymmetryforce1}
\end{equation}

\textit{Calculation of (C2)}\\


Here we cannot include  the tensor $(3\hat{r}_l\hat{r}_m-\delta_{lm})$ within the differentiation $\partial_x $ anymore: at the considered position, $\partial_x \hat{\boldsymbol{r}}(x,y,z) |_{x=z=0, y \neq0} \neq \boldsymbol{0}.$ Consequently, we evaluate each component separately. The only terms contributing to $\partial_i[...]$ (evaluated at $x=0$) are $G_{xy}=3xy/r^5=G_{yx}$, which present a linear term in $x$. Also, one has $G_{xz}=0=G_{zx}$ as the derivative is evaluated at $z=0$. Hence, 
 \begin{equation}
       \mathcal{G}_{xx}(y\boldsymbol{\hat{y}}) = \frac{2}{(4\pi\varepsilon_0)^2}\left(\partial_x\frac{3xy}{r^5}\right)_{\boldsymbol{r}=y\boldsymbol{\hat{y}}}^2 = \frac{18}{(4\pi\varepsilon_0)^2y^8} \, , \label{resultsymmetryforce1b}
 \end{equation}
 where the factor 2 comes from the fact that the terms $(l,m)=(x,y),(y,x)$ contributes equally.  Comparing eqs.(\ref{resultsymmetryforce1},\ref{resultsymmetryforce1b}) with eqs.(\ref{C1},\ref{C2}) we obtain $\kappa_1=36$ and $\kappa_2 = 18$, so that 
\begin{equation}
       \mathcal{G}_{ik}(\boldsymbol{r}) = \frac{18(2\hat{r}_i\hat{r}_k+\delta_{ik})}{(4\pi\varepsilon_0)^2r^8} \label{symmetryforce1}\, .
\end{equation}
Substituting in Eq.(\ref{d1cov}) we obtain
\begin{equation}
    \boldsymbol{F}^{(1)}=-\frac{9\hbar}{2(4\pi\varepsilon_0)^2}\left[\frac{2\boldsymbol{v}\cdot\boldsymbol{\hat{r}}}{r^8}\boldsymbol{\hat{r}}+\frac{\boldsymbol{v}}{r^8}\right]\Lambda^{(1)} \,.
\end{equation}

\vskip 0.3 cm

\textit{Calculation of $\mathcal{G}_{ikn}(\boldsymbol{r})$}

\vskip 0.3 cm

From the definition (\ref{defGcal}) and Eq.(\ref{E dipSM}) we have that
\begin{equation}\label{gcalik}
    \mathcal{G}_{ikn}(\boldsymbol{r}) = \left[\partial_k\partial_n\frac{(3\hat{r}_l\hat{r}_m-\delta_{lm})}{4\pi\varepsilon_0r^3} \right]\partial_i\frac{3\hat{r}_l\hat{r}_m-\delta_{lm}}{4\pi\varepsilon_0r^3}  \, .
\end{equation}
The most general three-rank tensor we can write is given by
\begin{equation}
   \mathcal{G}_{ikn}(\boldsymbol{r})  = \frac{\kappa_3\hat{r}_i\hat{r}_k\hat{r}_n+\kappa_4\delta_{kn}\hat{r}_i+\kappa_5(\delta_{ik}\hat{r}_n+\delta_{in}\hat{r}_k)}{(4\pi\varepsilon_0^2)r^9} \, ,
\end{equation}
where we used that the tensor is symmetric under the exchange $k\leftrightarrow n$. Since $G_{lm}$ represents the $l$-th component of the electrostatic field of a unit dipole pointing in the $\boldsymbol{e}_m$ direction, we have that $\boldsymbol{\nabla}^2 G_{lm}=0$ [this can be demonstrated noting that $\boldsymbol{\nabla}\times\boldsymbol{\nabla}\times\boldsymbol{E}=-\boldsymbol{\nabla}^2\boldsymbol{E}=\boldsymbol{0}$]. The contraction in components $k,n$ must therefore vanish, leaving us with the condition
\begin{equation}
    \kappa_3+3\kappa_4+2\kappa_5 =0 \, . \label{laplacian}
\end{equation}
We need in addition to consider two configurations which we choose to be: (C3) $i,k,n=x$ and $\boldsymbol{r}=x\boldsymbol{\hat{x}}$ and (C4) $i,k=y, n=x$ and $\boldsymbol{r}=x\boldsymbol{\hat{x}}$ for which we have
\begin{eqnarray}
   \mathcal{G}_{xxx}(x\boldsymbol{\hat{x}}) &=& \frac{\kappa_3+\kappa_4+2\kappa_5}{(4\pi\varepsilon_0)^2x^9} \;\; \mbox{(C3)} \, , \label{C3} \\
       \mathcal{G}_{yyx}(x\boldsymbol{\hat{x}}) &=& \frac{\kappa_5}{(4\pi\varepsilon_0)^2x^9} \;\; \mbox{(C4)} \, . \label{C4}
\end{eqnarray}
\textit{Calculation of (C3)}

For this configuration we can again contract the index $l$ with $m$ before performing the derivation:

\begin{equation}
     \mathcal{G}_{xxx}(x\boldsymbol{\hat{x}})  =\frac{6}{(4\pi\varepsilon_0)^2}\left(\partial_x\frac{1}{x^3}\right)\left(\partial_x^2\frac{1}{x^3}\right)=-\frac{216}{(4\pi\varepsilon_0)^2x^9} \, . 
\end{equation}

\textit{Calculation of (C4)}

The only contribution where $\partial_yG_{lm}$ does not vanish after taking $y=z=0$ is contained in $G_{xy}=G_{yx}$, as in configuration C2. We have that $\partial_y G_{xy}=1/x^4$ and $\partial_x\partial_y G_{xy}=\partial_x x^{-4}=-4/x^5$, so that
\begin{equation}
      \mathcal{G}_{yyx}(x\boldsymbol{\hat{x}})  = -\frac{8}{(4\pi\varepsilon_0)^2x^9} \, .
\end{equation}
Combining the results of C3 and C4 with Eq.~(\ref{laplacian}) we obtain $\kappa_3=-308,\kappa_4=108$ and $\kappa_5=-8$. We therefore obtain 
\begin{equation}
    (\partial_iG_{lm})(\partial_k\partial_n G_{lm}) = \frac{-308\hat{r}_i\hat{r}_k\hat{r}_n+108\delta_{kn}\hat{r}_i-8(\delta_{ik}\hat{r}_n+\delta_{in}\hat{r}_k)}{(4\pi\varepsilon_0)^2r^9} \, .
\end{equation}

\vskip 0.3 cm

\textit{Calculation of $\mathcal{G}_{iknp}(\boldsymbol{r})$}

\vskip 0.3 cm

From the definition (\ref{defGcal}) and Eq.(\ref{E dipSM}) we have that
\begin{equation}\label{gcalik}
    \mathcal{G}_{ikns}(\boldsymbol{r}) = \left[\partial_k\partial_n\partial_s\frac{(3\hat{r}_l\hat{r}_m-\delta_{lm})}{4\pi\varepsilon_0r^3} \right]\partial_i\frac{3\hat{r}_l\hat{r}_m-\delta_{lm}}{4\pi\varepsilon_0r^3}  \, .
\end{equation}
The most general four-rank tensor we can write is given by
\begin{equation} \label{ordem4sym}
    \mathcal{G}_{ikns}(\boldsymbol{r}) = \frac{\kappa_6\hat{r}_i\hat{r}_k\hat{r}_n\hat{r}_s+\kappa_7\hat r_{i}\delta_{\{kn}\hat{r}_{s\}}+\kappa_8\delta_{i\{k}\hat r_{n}\hat{r}_{s\}}+\kappa_9\delta_{\{ik}\delta_{ns\}}}{(4\pi\varepsilon_0)^2r^{10}} \, ,
\end{equation}
where $\{\cdots\}$ means that the index inside the bracket must be symmetrized. Let us consider the configurations (C5): $i,k,n,s=x$ and $\boldsymbol{r}=x\boldsymbol{\hat{x}}$  and  (C6): $i,k,n,s=x$ and $\boldsymbol{r}=y\boldsymbol{\hat{y}}$  
\begin{eqnarray}
      \mathcal{G}_{xxxx}(x\boldsymbol{\hat{x}}) &=& \frac{\kappa_6+3\kappa_7+3\kappa_8+3\kappa_9}{(4\pi\varepsilon_0)^2x^{10}} \;\; \mbox{(C5)} \, , \label{C5} \\
      \mathcal{G}_{xxxx}(y\boldsymbol{\hat{y}})  &=& \frac{3\kappa_9}{(4\pi\varepsilon_0)^2y^{10}} \;\; \mbox{(C6)} \, , \label{C6} 
\end{eqnarray}
\textit{Calculation of (C5)}

For this configuration we can again contract $l$ with $m$ before performing the derivation, leading us to

\begin{eqnarray}
   \mathcal{G}_{xxxx}(x\boldsymbol{\hat{x}}) &=&\frac{6}{(4\pi\varepsilon_0)^2}\left(\partial_x\frac{1}{x^3}\right)\left(\partial_x^3\frac{1}{x^3}\right)\cr\cr  
    &=&\frac{1080}{(4\pi\varepsilon_0)^2x^{10}} \, .  \label{k6k72}
\end{eqnarray}
\textit{Calculation of (C6)}

For the same reason as in C4 only $G_{xy}$ and $G_{yx}$ contribute. We have $\partial_x G_{xy}=y/r^5-5x^2y/r^7$. For the term involving only one derivative we evaluate this expression at $x=0$ and obtain $\partial_x G_{xy}=1/y^4$ as before. For the third derivative term we get: $\partial_x^3 G_{xy}=-5/y^6-10/y^6=-15/y^6$. Hence, taking into account a factor 2 coming from $G_{yx}$ we obtain

\begin{equation}
\mathcal{G}_{xxxx}(y\boldsymbol{\hat{y}})=  -\frac{30}{y^{10}} \, . \label{k9}
\end{equation}
As we saw, tensor (\ref{ordem4sym}) must vanish upon contracting $n$ with $s$. This contraction yields a second order tensor with components $\hat{r}_i\hat{r}_k$ and $\delta_{ik}$ and for it to be zero we obtain two conditions:
\begin{eqnarray}
    \kappa_6+5\kappa_7+2\kappa_8&=&0 \label{k6k71} \, ,\\ 
    \kappa_8+5\kappa_9 &=& 0 \, .
\end{eqnarray}
From Eq.~(\ref{k9}) we obtain $\kappa_9=-10$ and thus from previous equation we obtain $\kappa_8=50$. From eqs.(\ref{C5}), (\ref{k6k72}) and (\ref{k6k71}) we obtain the system of eqs.
\begin{eqnarray}
    \kappa_6+3\kappa_7&=&960 \, ,\\
    \kappa_6+5\kappa_7&=& -100 \, .
\end{eqnarray}
Solving it we obtain $\kappa_7=-530$ and $\kappa_6 = 2550$. Finally, we obtain
\begin{equation} \label{ordem4sym_calculado}
     \mathcal{G}_{ikns}(\boldsymbol{r}) = \frac{10(255\hat{r}_i\hat{r}_k\hat{r}_n\hat{r}_s-53\hat r_{i}\delta_{[kn}\hat{r}_{s]}+5\delta_{i[k}\hat r_{n}\hat{r}_{s]}-\delta_{[ik}\delta_{ns]})}{(4\pi\varepsilon_0)^2r^{10}} \, ,
\end{equation}
For a uniform motion ($\boldsymbol{a}=\boldsymbol{0}=\boldsymbol{j}$), equation (\ref{d3cov}) implies that 
\begin{eqnarray}
     \boldsymbol{F}^{(3)}&=&-\frac{5\hbar}{2(4\pi\varepsilon_0)^2}\Bigg[\frac{5(\boldsymbol{v}\cdot\boldsymbol{\hat{r}})^2-v^2}{r^{10}}\boldsymbol{v}\; +\cr\cr
     &+& \frac{85(\boldsymbol{v}\cdot\boldsymbol{\hat{r}})^3-53(\boldsymbol{v}\cdot\boldsymbol{\hat{r}})v^2}{r^{10}}\boldsymbol{\hat{r}}\Bigg]\Lambda^{(3)} \label{f3}
\end{eqnarray}

\section{Determination of the sign of $\Lambda^{(n)}$ for odd orders of $n$}

All the information regarding the internal atomic properties is contained in the function $\Lambda^{(n)}$, defined in the main text as 
\begin{equation}
    \Lambda^{(n)} =\frac{(-i)^n}{2\pi }\int_{-\infty}^{\infty}
    \left( \frac{d^n\alpha^A(\omega)}{d\omega^n}\eta^{B}(\omega)+A\longleftrightarrow B\right) d\omega \, . \label{lambdanfourier}
\end{equation}
 For $n=2k+1$, with $k$ a non-negative integer, we obtain 
\begin{equation}
    \Lambda^{(2k+1)} =\frac{(-1)^k}{\pi }\int_{0}^{\infty}
    \left( \frac{d^{2k+1}\alpha_I^{A}(\omega)}{d\omega^{2k+1}}\eta^{B}(\omega)+A\longleftrightarrow B\right) d\omega \, . 
\end{equation}
where we used that  $\eta(\omega)$ and $\alpha_R(\omega)=\mbox{Re}\,\alpha(\omega)$ are even functions while $\alpha_I(\omega)=\mbox{Im}\,\alpha(\omega)$ is an odd function. In the zero temperature limit, we have that $\eta^{B}(\omega)=2\,\mbox{sgn}\,(\omega)\alpha_I^B(\omega)$ simplifying the above expression to
\begin{equation}
    \Lambda^{(2k+1)} =\frac{(-1)^k}{\pi }\int_{0}^{\infty}
    \left( \frac{d^{2k+1}\alpha_I^{A}(\omega)}{d\omega^{2k+1}}\alpha_I^{B}(\omega)+A\longleftrightarrow B\right) d\omega \, . \label{lambda2k+1fourier}
\end{equation}
We know that $\alpha(\omega\to\infty)\to 0$ since the atom cannot respond to a high frequency incident field. Also, $\alpha_I(\omega)$ is an odd function, implying that $d^j\alpha/d\omega^j|_{\omega=0}=0$ for even $j$. After performing $2k+1$ integrations by parts we are left only with the surviving surface terms, leading us to
\begin{equation}
    \Lambda^{(2k+1)} =\frac{(-1)^k}{\pi } \sum_{j=0}^{k-1} \left(\frac{d^{(2k-2j-1)}\alpha_I^A(\omega)}{d\omega^{(2k-2j-1)}}\right)_{\omega=0}\left(\frac{d^{(2j+1)}\alpha_I^B(\omega)}{d\omega^{(2j+1)}}\right)_{\omega=0} \, , \label{lambda2k+1}
\end{equation}
valid for $k\geq 1$. For $k=0$ we have $\Lambda^{(1)}=0$. 

To proceed we must choose a model for the atomic polarizability. In the absence of dissipation, the Fourier transform of Eq. (\ref{alpha}) yields \cite{Santos24} 
\begin{equation}
    \alpha^A(\omega) = \sum_a\frac{\omega_a\alpha_{a}(0)}{2}\left[\frac{1}{\omega+\omega_a}-\frac{1}{\omega-\omega_a}\right] \, , \label{alphamultilevelsemdissipacao}
\end{equation}
where $\hbar\omega_a$ denote the energy of the internal atomic states and $\alpha_a(0)\geq 0$ accounts for the contribution of transition $\omega_a$ for the static polarizability. Naturally, in the absence of dissipation $\alpha_I^A=0$ and there is no quantum friction. Dissipation can be included phenomenologically by taking into account a finite width on each transition frequency:
\begin{equation}
    \alpha^A(\omega) = \sum_a\frac{\omega_a\alpha_{a}(0)}{2}\left[\frac{1}{\omega+\omega_a+i\gamma_a/2}-\frac{1}{\omega-\omega_a+i\gamma_a/2}\right] \, , \label{alphamultilevel}
\end{equation}
where $\gamma_a$ are positive damping coefficients. We use here that $\gamma_a\ll\omega_a$, which is typically the case. This model is in accordance with causality since the poles $\omega^{*}$ have $\mbox{Im}\,\omega^*<0$. In addition, $\mbox{Im}\,\alpha^A(\omega)$ ($\mbox{Re}\,\alpha^A(\omega)$) is an odd (even) function, in accordance with the reality of $\alpha^A(\tau)$.  

In order to obtain $\Lambda^{(2k+1)}$ we see from Eq.(\ref{lambda2k+1}) that we need to calculate the derivatives of odd order of $\alpha(\omega)$ evaluated at $\omega=0$. In the low dissipative case ($\lambda_a\ll \omega_a$) this can be readily done noticing that 
\begin{eqnarray}
    \frac{d^n\alpha(\omega)}{d\omega^n}\Bigg|_{\omega=0}&=&\sum_{a}\frac{\omega_a\alpha_a(0)}{2}\frac{d^n}{d\omega_a^n}\left[\frac{1}{\omega+\omega_a+i\gamma_a/2}+\frac{1}{\omega-\omega_a+i\gamma_a/2}\right]_{\omega=0} \cr\cr
    &\approx&\sum_{a}\frac{\omega_a\alpha_a(0)}{2}\frac{d^n}{d\omega_a^n}\frac{(-i)\gamma_a}{\omega_a^2} \, .
\end{eqnarray}
Therefore,
\begin{equation}
    \frac{d^{(2k+1)}\alpha_I(\omega)}{d\omega^{(2k+1)}}\Bigg|_{\omega=0} \approx (2k+2)!\sum_{a}\frac{\alpha_a(0)}{2}\frac{\gamma_a}{\omega_a^{(2k+2)}} \, , \label{dalphadomegazero}
\end{equation}
which is positive for every $k$. A similar analysis holds for the atom $B$. Therefore, from Eq.(\ref{lambda2k+1}) we see that $\Lambda^{(2k+1)}$ has the sign of $(-1)^k$.  

Evaluating explicitly the case $k=1$ we obtain from Eq.(\ref{lambda2k+1})

\begin{equation}
    \Lambda^{(3)}=-\frac{1}{\pi}\left(\frac{d}{d\omega}\alpha_I^{A}\right)_{\omega=0}\left(\frac{d}{d\omega}\alpha_I^{B}\right)_{\omega=0} \, . \label{lambda3geral}
\end{equation}
Substituting Eq.(\ref{dalphadomegazero}) and an analogous relation to atom $B$ into the previous equation we obtain 
\begin{equation}
    \Lambda^{(3)}= -\sum_{a,b} \frac{\alpha_{a}(0)\alpha_{b}(0)\gamma_{a}\gamma_{b}}{\pi\omega_{a}^2\omega_{b}^2} \, . \label{lambda2lor}
\end{equation}
For the sake of completeness, we include the expression for $\Lambda^{(2k+1)}$:
\begin{equation}
    \Lambda^{(2k+1)} =(-1)^k \sum_{a,b}\frac{\alpha_a(0)\alpha_b(0)\gamma_a\gamma_b}{4\pi}\sum_{j=0}^{k-1} \frac{(2k-2j)!(2j+2)!}{\omega_a^{(2k-2j)}\omega_b^{(2j+2)}} \, , \label{lambda2k+1_result}
\end{equation}
valid for $k\geq 1$.

\section{Orientation average of the force} 

In this section we detail the calculation regarding the connection of microscopic quantum friction with a macroscopic situation. We analyze the force exerted by a rarefied medium on an atom moving with constant velocity $\boldsymbol{v}$ parallel to its surface. The rarefied assumption allows to evaluate the force by a pairwise superposition of the force between the moving atom and each atom in the medium. Denoting by $\boldsymbol{F}_{\rm net}^{(n)}$ the $n$-th dynamical correction to the total force and by $\boldsymbol{F}^{(n)}$ the $n$-th dynamical correction to the force between two atoms we have that
\begin{equation}
    \boldsymbol{F}_{\rm net}^{(n)} =2\pi N\int_{-\infty}^0 dz'\int_0^{\infty}d\rho \rho \langle \boldsymbol{F}^{(n)}(\boldsymbol{\rho}+(z_0-z')\boldsymbol{\hat{z}} )\rangle \, , \label{fnet}
\end{equation}
where
\begin{equation}
    \langle \mathcal{O}\rangle = (2\pi)^{-1}\int_0^{2\pi} \mathcal{O}(\phi)d\phi \, .
\end{equation}
Symmetry and dimensional considerations guarantee that for even $n$, denoted by $n=2k$, we have
\begin{eqnarray}
    \boldsymbol{F}^{(2k)}(\boldsymbol{r})&=&\frac{\hbar\Lambda^{(2k)}}{(4\pi\varepsilon_0)^2}\Bigg[\sum_{j=0}^{k}  c_{2j}^{(2k)}\frac{\boldsymbol{v}^{2(k-j)}(\boldsymbol{v}\cdot\boldsymbol{r})^{2j}\boldsymbol{r}}{r^{2(j+k+4)}} +\cr\cr 
    &+&\sum_{j=0}^{k-1}  c_{2j+1}^{(2k)}\frac{\boldsymbol{v}^{2(k-j-1)}(\boldsymbol{v}\cdot\boldsymbol{r})^{2j+1}\boldsymbol{v}}{r^{2(j+k+4)}}\Bigg]
    \, ,
\end{eqnarray}
where $c_{m}^{(2k)}$ are dimensionless coefficients. For $k=0$ only the first term is present. For motion parallel to the plate, we have that $\boldsymbol{v}\cdot\boldsymbol{r}=\boldsymbol{v}\cdot\boldsymbol{\rho}$.  Denoting by $\rho_i$ the i-th cartesian component of $\boldsymbol{\rho}$  we have that the average $\langle \rho_{i_1}\cdots\rho_{i_m}\rangle$ vanishes by symmetry when $m$ is odd. Therefore, writing $\boldsymbol{r}=\boldsymbol{\rho}+z_0\boldsymbol{\hat{z}}$, we see that 
\begin{eqnarray}
 &&    \langle \boldsymbol{F}^{(2k)}(\boldsymbol{\rho}+z_0\boldsymbol{\hat{z}}) \rangle  = \frac{\hbar\Lambda^{(2k)}}{(4\pi\varepsilon_0)^2}\cr\cr 
 &&\sum_{j=0}^{k}  c_{2j}^{(2k)}\frac{\boldsymbol{v}^{2(k-j)}(v_{i_1}\cdots v_{i_{2j}})\langle \rho_{i_1}\cdots\rho_{i_{2j}}\rangle z_0}{(\rho^2+z_0^2)^{(j+k+4)}} \boldsymbol{\hat{z}}
\end{eqnarray}
Therefore, we demonstrate on general grounds that dynamical corrections of the force with even $n$ produce only a correction on the normal component of the force exerted by the plate. There is no lateral force and therefore no influence in this case to the quantum friction, as expected by the discussion on the previous section. Note that this result does not require thermal equilibrium since we did not make any assumption on $\Lambda^{(2k)}$. 

For odd $n$, denoted by $n=2k+1$, symmetry and dimensional arguments show that the force must have the form
\begin{eqnarray}
    \boldsymbol{F}^{(2k+1)}(\boldsymbol{r})&=&\frac{\hbar\Lambda^{(2k+1)}}{(4\pi\varepsilon_0)^2}\Bigg[\sum_{j=0}^{k}  c_{2j}^{(2k+1)}\frac{\boldsymbol{v}^{2(k-j)}(\boldsymbol{v}\cdot\boldsymbol{r})^{2j+1}\boldsymbol{r}}{r^{2(j+k+5)}} +\cr\cr 
    &+&\sum_{j=0}^{k}  c_{2j+1}^{(2k+1)}\frac{\boldsymbol{v}^{2(k-j)}(\boldsymbol{v}\cdot\boldsymbol{r})^{2j}\boldsymbol{v}}{r^{2(j+k+4)}}\Bigg]
    \, ,
\end{eqnarray}
where $c_{m}^{(2k)}$ are dimensionless coefficients. The cases $k=0$ and $k=1$ were obtained, respectively, in Eqs.(\ref{f1}) and (\ref{f3}). After performing the angular average we see that the force is parallel to the plate, in contrast to what happened with $n$ even. Let us now evaluate explicitly the $n=1$ and $n=3$ cases.

We demonstrated in the main text that
\begin{equation}\label{f1}
     \boldsymbol{F}^{(1)}(\boldsymbol{r})=-\frac{9\hbar}{2(4\pi\varepsilon_0)^2}\left[\frac{2\boldsymbol{v}\cdot\boldsymbol{r}}{r^{10}}\boldsymbol{r}+\frac{\boldsymbol{v}}{r^8}\right]\Lambda^{(1)} \,. 
\end{equation}
The assumption that the atom moves parallel to the plate allows for the substitution of $\boldsymbol{v}\cdot\boldsymbol{r}$
by $\boldsymbol{v}\cdot\boldsymbol{\rho}$. We have that the $i$-th component of the angular averaged force is given by  
\begin{equation}
    \langle F^{(1)}_i(\boldsymbol{\rho}+z_0\boldsymbol{\hat{z}} )\rangle =-\frac{9\hbar\Lambda^{(1)}}{2(4\pi\varepsilon_0)^2}\left[\frac{2v_j\langle \rho_jr_i\rangle}{(\rho^2+z_0^2)^5}+\frac{v_i}{(\rho^2+z_0^2)^4}\right]
 \end{equation}
Note that $\langle \rho_jz\rangle = z\langle \rho_j\rangle = 0$ and 
\begin{equation}
\langle \rho_j\rho_i\rangle = \frac{\rho^2}{2}\delta_{ij} \,\,\,(i,j=x,y)    \, . \label{averagerhoirhoj}
\end{equation}
Hence,
\begin{equation}
     \langle \boldsymbol{F}^{(1)}(\boldsymbol{\rho}+z_0\boldsymbol{\hat{z}} )\rangle =-\frac{9\hbar\Lambda^{(1)}}{(4\pi\varepsilon_0)^2}\frac{\boldsymbol{v}}{(\rho^2+z_0^2)^4}\, .
\end{equation}
Substituting in Eq.(\ref{fnet}) we obtain
\begin{equation}
    \boldsymbol{F}_{\rm net}^{(1)}=-\frac{3n\hbar\Lambda^{(1)}}{16\pi\varepsilon_0^2}\frac{\boldsymbol{v}}{z_0^6} \, .
\end{equation}
Let us now evaluate the $n=3$ dynamical correction. As demonstrated in the main text
\begin{equation}\label{F3completo}
    \boldsymbol{F}^{(3)}=-\frac{5\hbar\Lambda^{(3)}}{2(4\pi\varepsilon_0)^2}\Bigg[\frac{5(\boldsymbol{v}\cdot\boldsymbol{r})^2}{r^{12}}\boldsymbol{v}-\frac{v^2}{r^{10}}\boldsymbol{v}+ \frac{85(\boldsymbol{v}\cdot\boldsymbol{r})^3}{r^{14}}\boldsymbol{r}-\frac{53(\boldsymbol{v}\cdot\boldsymbol{r})v^2}{r^{12}}\boldsymbol{r}\Bigg] \, .
\end{equation}
For motion parallel to the plate we have $\boldsymbol{v}\cdot\boldsymbol{r}=\boldsymbol{v}\cdot\boldsymbol{\rho}$. Let us analyze the angular average term by term. From Eq.(\ref{averagerhoirhoj}) we have that
\begin{eqnarray}
    \langle (\boldsymbol{v}\cdot\boldsymbol{r})^2\rangle&=&\frac{\rho^2v^2}{2} \, ,\\
    \langle (\boldsymbol{v}\cdot\boldsymbol{r})\boldsymbol{r}\rangle&=&\frac{\rho^2\boldsymbol{v}}{2} \, ,
\end{eqnarray}
where in the last equality we used that $\langle (\boldsymbol{v}\cdot\boldsymbol{\rho})\boldsymbol{\hat{z}\rangle} =\boldsymbol{0}$. We also have that $\langle (\boldsymbol{v}\cdot\boldsymbol{\rho})\boldsymbol{\hat{z}}\rangle=\boldsymbol{0}$ and that
\begin{equation}
    \langle \rho_{i}\rho_j\rho_k\rho_l\rangle=\frac{\rho^4}{8}(\delta_{ij}\delta_{kl}+\delta_{ik}\delta_{jl}+\delta_{il}\delta_{jk})  \,\,\,(i,j,k,l=x,y)  \, .
\end{equation}
Therefore, 
\begin{equation}
    \langle(\boldsymbol{v}\cdot\boldsymbol{\rho})^3\boldsymbol{r}\rangle=\frac{3v^2\boldsymbol{v}\rho^4}{8} \, .
\end{equation}
Taking the angular average of Eq.(\ref{F3completo})
\begin{equation}
    \langle \boldsymbol{F}^{(3)}(\boldsymbol{r})\rangle =-\frac{5\hbar\Lambda^{(3)}v^2\boldsymbol{v}}{2(4\pi\varepsilon_0)^2}\left[\frac{255\rho^4}{8r^{14}}-\frac{24\rho^2}{r^{12}}-\frac{1}{r^{10}}\right]\, .
\end{equation}
Writing $\rho^2=r^2-z_0^2$ we obtain
\begin{equation}
    \langle \boldsymbol{F}^{(3)}(\boldsymbol{\rho}+z_0\boldsymbol{\hat{z}})\rangle =-\frac{5\hbar\Lambda^{(3)}v^2\boldsymbol{v}}{2(4\pi\varepsilon_0)^2}\left[\frac{255z_{0}^4}{8(\rho^2+z_0^2)^{7}}-\frac{159z_0^2}{4(\rho^2+z_0^2)^{6}}+\frac{55}{8(\rho^2+z_0^2)^{5}}\right]\, .
\end{equation}
 Performing the integration given in Eq.(\ref{fnet}) we obtain
\begin{equation}
    \boldsymbol{F}_{\rm net}^{(3)} = \frac{147n\pi\hbar\Lambda^{(3)}v^2}{64(4\pi\varepsilon_0)^2z_0^8}\boldsymbol{v} \, .
\end{equation}